\documentstyle[aps,12pt]{revtex}
\begin{document}

\newcommand{\nc}{\newcommand}
\newcommand{\BE}{\begin{equation}}
\newcommand{\EE}{\end{equation}}
\newcommand{\BA}{\begin{eqnarray}}
\newcommand{\EA}{\end{eqnarray}}

\title{Effects of Inelastic Scattering on Tunneling Time
\\in Generalized Nelson's Quantum Mechanics}

\author {\vspace{0.3cm}\\Kentaro Imafuku$^{1}$\footnote{E-mail
696L5051@cfi.waseda.ac.jp},  Ichiro Ohba$^{1,2,3}$\footnote{E-mail
ohba@cfi.waseda.ac.jp}, and  Yoshiya
Yamanaka$^{4}$\footnote{E-mail yamanaka@cfi.waseda.ac.jp}\vspace{0.5cm}}

\address{Department of Physics$^{1}$,\\
Advanced Research Center for Science and Engineering$^{2}$,\\
Kagami Memorial Laboratory for Materials Science
and Technology$^{3}$,\\ Waseda University Shinjuku-ku, Tokyo 169, Japan\\
Waseda University Senior High School$^{4}$\\
Nerima-ku, Tokyo 177, Japan}
\date{\today}

\vspace{2cm}
\maketitle
\vspace{2cm}

\begin{abstract}
We analyze the effects of inelastic scattering on the
tunneling time theoretically, using generalized Nelson's
quantum mechanics. This generalization enables us to
describe quantum system with optical potential and channel
couplings in a real time stochastic approach, which seems
to give us a new insight into quantum mechanics beyond
Copenhagen interpretation.
\end{abstract}
\newpage

\section{Introduction}

An issue of the tunneling time, i.e., the time
associated with the passage of a particle through
a tunneling barrier, has been discussed in many theoretical
studies  \cite{Wigner=1955,Buttiker=1982,Lee=1983,Buttiker=1983,Sokolvski=1987,Fertig=1990,Chen=1990,Sokolovski=1991,Olkhovsky=1992,Leavens=1993BK,Martin=1993,Leavens=1994,Brouard=1994,Steinberg=1995PRL,Collins=1987,Hauge=1989,Landauer=1994} and is not settled yet.
This difficulty arises mainly from the fact that time is
not an observable represented by a self-adjoint operator
but just a parameter in quantum mechanics.

In our previous paper \cite{Imafuku=1995}, we proposed
a new method to evaluate the tunneling time, using
Nelson's approach of quantum mechanics \cite{Nelson=1966}.
Our aim then was to treat tunneling effects in a detailed
time-dependent and fully quantum mechanical way, as any
theoretical expression of the tunneling time must be tested
by experiments which are feasible at present and in near future.

As discussed in the reference \cite{Imafuku=1995}, the
Nelson's approach of quantum mechanics has several advantages
to study the tunneling time, a few of which are listed below.

First of all, this approach using the real time stochastic process
enables us to describe individual experimental runs of a quantum system
in terminology of ``analogue'' of classical mechanics.  This
is true even in the
tunnel region where classical path is forbidden.  From sample paths
generated by the stochastic process we obtain information on the time
parameter, in particular, the tunneling time.

As a matter of course, the whole ensemble of sample paths gives
us all the same results as quantum mechanics in ordinary approach does,
e.g., expectation values
of observable, transmission and reflection probabilities in scattering
problem and so on.  It is important for us to note that in scattering
phenomena (those without bound states)
the transmission and reflection ensembles are
defined unambiguously, namely each sample path is classified distinctively
into either transmission ensemble or reflection one.

We need to accumulate a sufficient
number of sample paths in numerical simulations.
In thick or/and high potential case the transmission
probability is low and consequently we have a difficulty
that a number of sample paths belonging to the transmission
ensemble is also low when each sample path is followed in
forward time direction.  However, in Nelson's approach
there is not only the forward Langevin equation but also
the backward Langevin equation  (see (\ref{Ito_b}) below),
both being equivalent to each other in physical results.
The difficulty above is avoided when the backward
Langevin equation is employed.

Taking account of these advantages, we have developed
a theoretical formulation  of time-dependent description
of tunneling phenomena based on the Nelson's stochastic approach
in \cite{Imafuku=1995}.  Numerical simulations
for a one-dimensional square well potential barrier model
were demonstrated.  An important result about the tunneling time
then is that there are the three characteristic
times, i.e., {\em the passing time and the hesitating time}, 
and their sum, {\em the interacting time}.  The probability 
distribution of these three times were calculated numerically.

Our previous study treated only a quantum system
of a single particle under a simple potential.  But realistic
experimental situations are much complicated. Naturally we are
tempted to extend our previous formulation to more general
scattering phenomena.  In this paper we consider cases in which
transition processes into other channels or absorptive processes
takes place during scattering processes, and look into these effects on
the tunneling time.

Processes of transition into other channels and absorption
are described by  channel coupling and optical potential (complex potential),
respectively, in the ordinary quantum mechanics using the
Schr\"{o}dinger equation.  So far
it is known  that the Nelson's formulation is
equivalent to the Schr\"{o}dinger equation {\em only for a one-body problem
with a single channel and a real potential}.
The purpose of this paper is to generalize the Nelson's
stochastic quantization so that it can deal with multi-channel coupling and/or
optical potential problems.  As will be shown below, one can construct
such generalized formulations of the Nelson's approach with additional 
stochastic jumping processes. 
These theoretical formulations allow us to perform numerical simulations
of stochastic processes as before \cite{Imafuku=1995}. 
This way we can investigate the effects of transition into other channels
or absorption on the tunneling time.

The paper is organized as follows:  In the next section  the 
original Nelson's quantum mechanics is reviewed briefly for later relevance.
We propose a formulation of the Nelson's approach, generalized to a
quantum system with channeling-coupling in Sec.~3.
The formulation
of Sec.~3 hints us how to develop a formulation for optical potential,
which is shown in Sec.~4.  In Sec.~5 numerical simulation for square well 
potential model, using the formulations in 
Sections 3 and 4, are demonstrated, and physical implications of these results
are analyzed.  The final section is devoted to summary and some comments.

\section{Brief review of Nelson's quantum mechanics}

We start with a brief review of the original Nelson's quantum
mechanics which consists of two basic conditions, i.e., the kinematical
condition and the dynamical one.

The kinematical condition is given by the
Ito-type stochastic differential equation:  There are two ways to express it,
depending on forward or backward time direction.
Explicitly we have for forward time evolution,
\BE \label{Ito_f}
dx(t)=b(x(t),t)dt+dw(t),
\EE
and for backward time evolution,
\BE \label{Ito_b}
dx(t)=b_{*}(x(t),t)dt+dw_{*}(t) .
\EE
The $dw(t)$ is the Gaussian white noise
(representing the quantum fluctuation) with the statistical
properties of
\BE \label{dw}
<dw(t)>=0 \qquad \mbox{and} \qquad
<dw(t) dw(t)>=\frac{\hbar}{m}dt,
\EE
and the same properties for $dw_{*}(t)$ as in (\ref{dw}).
Here $<\cdots>$ means a sample average.
It is easy to show that for these two Langevin equations hold the following
Fokker-Planck equations for the
distribution function $P(x,t)$ of the random variables
$x(t)$,
\BA \label{fokker_f}
\frac{\partial P(x,t)}{\partial t}
&=& \left[-\frac{\partial}{\partial x}b(x,t)
+\frac{\hbar}{2m}\frac{\partial^{2}}{\partial x^{2}}\right]P(x,t)
\quad \mbox{(forward in $t$)},\\
\label{fokker_b}
-\frac{\partial P(x,t)}{\partial t}
&=&\left[\frac{\partial}{\partial x}b_{*}(x,t)
+\frac{\hbar}{2m}\frac{\partial^{2}}{\partial x^{2}}\right]P(x,t)
\quad \mbox{(backward in $t$)} .
\EA
Thus a pair of equations (\ref{Ito_f}) and (\ref{Ito_b})
is mathematically equivalent to a pair of equations
(\ref{fokker_f}) and (\ref{fokker_b}).
We get an osmotic velocity, $u$ from the sum of Eqs.
(\ref{fokker_f}) and (\ref{fokker_b}) as
\begin{equation}\label{u_}
u=\frac{b-b_{*}}{2}=\frac{\hbar}{2m}\frac{1}{P}
\frac{\partial P}{\partial x}
\end{equation}
under the boundary condition of
\BE
P(x \rightarrow \infty,t)\rightarrow 0.
\EE
The subtraction (\ref{fokker_b}) from (\ref{fokker_f})
gives
\begin{equation}\label{v_}
\frac{\partial P}{\partial t}=
-\frac{\partial}{\partial x}(vP)
\end{equation}
where $v$ is a current velocity
\begin{equation}\label{defv}
v=\frac{b+b_{*}}{2} \ .
\end{equation}
The elimination of $P(x,t)$ from
(\ref{u_}) and (\ref{v_}) leads to an equation called
the kinematical equation,
\begin{equation}\label{kinematics}
\frac{\partial u}{\partial t}=
-\frac{\hbar}{2m}\frac{\partial ^{2} v}{\partial x^{2}}
-\frac{\partial}{\partial x}(uv).
\end{equation}

The dynamical condition is expressed through
the ``mean time derivatives" introduced as follows:
The ``mean forward time derivative" $Df(t)$ is defined as
\begin{equation} \label{MDF}
Df(t)\equiv \lim_{\Delta t \rightarrow +0}
< \frac{f(t+\Delta t)-f(t)}{\Delta t}|f(s)
\, (s \leq t)\, \mbox{fixed}>,
\end{equation}
and the ``mean backward time derivative" $D_{*}f(t)$ is defined as
\begin{equation} \label{MDB}
D_{*}f(t)\equiv \lim_{\Delta t \rightarrow +0}
<\frac{f(t)-f(t-\Delta t)}{\Delta t}| f(s)
\, (s\geq t) \, \mbox{fixed}>.
\end{equation}
The ``mean balanced acceleration" is introduced through the definitions of
(\ref{MDF}) and (\ref{MDB}) as
\begin{equation}
a(x(t),t)\equiv \frac{DD_{*}+D_{*}D}{2}x(t) .
\end{equation}
Note that this definition can be rewritten as
\begin{equation} \label{eqa}
a(x,t)=-\frac{\hbar}{2m}\frac{\partial^{2} u}{\partial x^{2}}
+\frac{1}{2}\frac{\partial}{\partial x}(v^{2}-u^{2})
+\frac{\partial v}{\partial t}
\end{equation}
from (\ref{Ito_f}) and (\ref{Ito_b}) with (\ref{u_}) and (\ref{defv}).
The dynamical condition is nothing but the classical Newton equation to this
``mean balanced acceleration" $a(x(t),t)$, that is,
\begin{equation}
m a(x,t) = - \frac{\partial V}{\partial x},
\end{equation}
from which we derive the ``Newton-Nelson equation",
\begin{equation}\label{Newton-Nelson}
\frac{\partial v}{\partial t}
=\frac{\hbar}{2m}\frac{\partial^{2} u}{\partial x^{2}}
-v \frac{\partial v}{\partial x}+u \frac{\partial u}{\partial x}
-\frac{1}{m}\frac{\partial V}{\partial x}
\end{equation}
because of (\ref{eqa}).

Summarize the mathematical structure of Nelson's quantum mechanics.
The two basic equations, (\ref{kinematics}) 
from the kinematical condition, and
(\ref{Newton-Nelson}) from the dynamical
condition, form a set of simultaneous equations for two unknown
functions $u(x,t)$ and $v(x,t)$, or equivalently $b(x,t)$
and  $b_{*}(x,t)$.
Then we can determined the ensemble of sample paths or the
distribution function $P(x,t)$.
Although it is practically very difficult to solve these equations directly
due to their nonlinearity, one can easily show the equivalence between this approach
and the ordinary approach of the Schr\"{o}dinger equation.
This fact 
tells us that one can solve the problem by means of the wave function
much more easily.
The equation
\BE\label{S_EQ}
\frac{\partial}{\partial x}\left[i \frac{\hbar}{m}\frac{1}{\psi '}
\frac{\partial \psi '}{\partial t}+\frac{1}{2} \left(\frac{\hbar}{m}\right)^{2}
\frac{1}{\psi '}\frac{\partial^{2}\psi '}{\partial x^{2}}-\frac{1}{m}V
\right]
=0
\EE
follows from the combination of
(\ref{kinematics})$+i$(\ref{Newton-Nelson}), where
\BE\label{u+iv}
u+iv=\frac{\hbar}{m}\frac{1}{\psi '}\frac{\partial \psi '}{\partial x}.
\EE
Equation (\ref{S_EQ}) clearly shows
the relationship between $\psi '$ and
the wave function $\psi$ as the solution of Schr\"{o}dinger equation
\begin{equation}\label{schrodinger}
i\hbar\frac{\partial \psi}{\partial t}
=\left(-\frac{\hbar^{2}}{2m}\frac{\partial^{2}}{\partial
x^{2}}+V \right)\psi,
\end{equation}
namely
\BE
\psi(x,t) =\psi'(x,t) \exp(-\frac{im}{\hbar}\int^{t}\eta(s) ds)
\EE
with an arbitrary function of $t$, $\eta(t)$, which has no physical relevance.
It is easily seen from this proof of the equivalence that one has
the expressions for $b(x,t)$, $b_{*}(x,t)$ and $P(x,t)$ in terms of
$\psi(x,t)$,
\BA\label{b}
b(x,t)&=&\frac{\hbar}{m}
({\rm Im+Re})\frac{\partial}{\partial x} \ln \psi(x,t), \\
\label{b*}
b_{*}(x,t)&=&\frac{\hbar}{m}
({\rm Im-Re})\frac{\partial}{\partial x} \ln \psi(x,t), \\
\label{rho}
P(x,t)& =&|\psi(x,t)|^{2}.
\EA

\section{Stochastic formulation for quantum system with channeling coupling}

We now turn to generalize the above Nelson's approach to
a system with a channel coupling.  For simplicity, consider 
the $2$-channel Schr\"{o}dinger
equations ($\{i,j\}=\{1,2\}$)
\begin{equation}\label{2chan}
i\hbar\frac{\partial }{\partial t}\psi_{i}(x,t)
=\left(-\frac{\hbar^{2}}{2m_{i}}\frac{\partial^{2}}{\partial x^{2}}
 +V_{ii}(x,t)\right)
\psi_{i}(x,t)+V_{ij}(x,t)\psi_{j}(x,t),
\end{equation}
with 
\begin{equation}
V_{ij}=V_{ji}^{*}.
\end{equation}
Here and below the dummy index does not imply taking a sum. 
As will be seen, the generalization of the formulation in this section 
to N-channel case (N$>2$) is straightforward.

Consider the Fokker-Planck equations in the stochastic formulation,
corresponding to (\ref{2chan}).  First we require a natural extension of
(\ref{rho}) to the present case,
\BE \label{ch_rho}
P_i(x,t) =|\psi_i(x,t)|^{2} .
\EE
The diagonal parts (the kinetic energy and $V_{ii}$ terms) in (\ref{2chan}) are
expected to be dealt with as in the previous section.  The Schr\"odinger
equations (\ref{2chan}) and their complex conjugates suggest the following
equations for $P_i(x,t)$,
\BE\label{ch_fokker_f}
\frac{\partial P_{i}(x,t)}{\partial t}
=\left[-\frac{\partial}{\partial x}b_i(x,t)+
\frac{\hbar}{2m_{i}}\frac{\partial^{2}}{\partial x^{2}}
-W_{(i\rightarrow j)}(x,t)\right]P_{i}(x,t) \qquad\mbox{(forward in time)},
\EE
\BE\label{ch_fokker_b}
-\frac{\partial P_{i}(x,t)}{\partial t}
= \left [\frac{\partial}{\partial x}b_{*i}(x,t)+
\frac{\hbar}{2m_{i}}\frac{\partial^{2}}{\partial x^{2}}
+W_{(i\rightarrow j)}(x,t)\right]P_{i}(x,t) \qquad \mbox{(backward in time)},
\EE
as $P_i(x,t)$ increases or decreases, due to the potential $V_{ij}$ causing
transitions between $i$ and $j$,
at the rate of the absolute value of
\BE \label{itoj}
W_{(i\rightarrow j)}P_{i}=-W_{(j\rightarrow i)}P_{j} = \frac{2}{\hbar}{\rm Im}
\psi^{*}_{j}V_{ji}\psi_{i}
\EE
Although the sum of (\ref{ch_fokker_f}) and (\ref{ch_fokker_b})
leads to (\ref{u_}) with the index $i$, 
\BE
\label{ch_u_}
u_i=\frac{b_i-b_{*i}}{2}=\frac{\hbar}{2m_i}\frac{1}{P_i}
\frac{\partial P_i}{\partial x} ,
\EE
their difference provides us with
\begin{equation}\label{ch_v_}
\frac{\partial P_i}{\partial t}=
-\frac{\partial}{\partial x}(v_i P_i) -W_{(i\rightarrow j)} P_i
\end{equation}
instead of (\ref{v_}),
where
\BE
\label{ch_defv}
v_i=\frac{b_i+b_{*i}}{2}  .
\EE
As a result, eliminating $P_i(x,t)$ from (\ref{ch_u_}) and (\ref{ch_v_}),
one derives the following kinematical equation
\begin{equation}\label{ch_kinematics}
\frac{\partial u_i}{\partial t}=
-\frac{\hbar}{2m_{i}}\frac{\partial ^{2} v_i}{\partial x^{2}}
-\frac{\partial}{\partial x}(u_i v_i)-\frac{\hbar}{2m_i}
\frac{\partial}{\partial x} W_{(i\rightarrow j)}
\end{equation}
instead of (\ref{kinematics}).

Here arises a natural question what are stochastic differential
equations corresponding to the Fokker-Planck equations in
(\ref{ch_fokker_f}) and (\ref{ch_fokker_b}), just as
(\ref{Ito_f}) and (\ref{Ito_b}) correspond to
(\ref{fokker_f}) and (\ref{fokker_b}).
Apparently we need two random variables $x_i(t)$ ($i=1,2$),
which are assumed to be 
subject to the stochastic differential equations, similar to (\ref{Ito_f}) 
and (\ref{Ito_b}),
\begin{eqnarray}
\label{ch_llf}
dx_i(t)&=&b_i(x_i(t),t)dt+dw_i(t) \qquad \mbox{(forward in time)},\\
\label{ch_llb}
dx_i(t)&=&b_{*i}(x_i(t),t)dt+dw_{*i}(t)
\qquad \mbox{(backward in time)},
\end{eqnarray}
with the properties for $dw_i(t)$ and $dw_{*i}(t)$,
\BA
<dw_i(t)>=0,&\qquad&
<dw_i(t)dw_j(t)>= \frac{\hbar}{m_{i}}\delta_{ij}dt  \nonumber \\
<dw_{*i}(t)>=0,&\qquad&
<dw_{*i}(t)dw_{*j}(t)>= \frac{\hbar}{m_{i}}\delta_{ij}dt . 
\label{dwidwj}
\EA
As is easily seen, a naive interpretation of 
these independent stochastic differential equations 
leads only to the 
Fokker-Planck equations in (\ref{ch_fokker_f}) and (\ref{ch_fokker_b}) 
{\em without the terms proportional to $W_{(i\rightarrow j)}$}.
An additional mechanism to take account of the quantum jump between $i$ and 
$j$ represented by the terms involving $W_{(i\rightarrow j)}$ is necessary.
For this purpose we supplement (\ref{ch_llf}) and (\ref{ch_llb}) with 
a stochastic jumping process between $i$ and $j$.  Thus we attempt
below the formulation for the two random variables $x_i(t)$, 
subject to the stochastic differential
equations (\ref{ch_llf}) and (\ref{ch_llb})
combined with a stochastic jumping process
in the following way.

The ``dynamical" rule to determine how each sample path $x_i(t)$
changes its index ($i=1\rightarrow  2$ or vice versa)
during passage of time 
is described by the following random jumping process (Fig. 1):  At each time
a dice is cast, {\em independently of the stochastic equation} (\ref{ch_llf})
and (\ref{ch_llb}), and each sample path either keeps or changes its index
at a certain rate.  
For forward time direction, we have the rule 
in case of $W_{(i\rightarrow j)}>0$
($i\neq j$),
\BA
x_i(t) &\longrightarrow&
\left\{
\begin{array}{l}
x_j(t+dt) \quad \mbox{with the probability of 
$W_{(i\rightarrow j)}(x_i(t),t)dt$} ,\\ 
x_i(t+dt) \quad \mbox{with the probability of 
$1-W_{(i\rightarrow j)}(x_i(t),t)dt$}
\end{array}  \right. , \nonumber \\
x_j(t) &\longrightarrow& \quad 
x_j(t+dt) \quad \mbox{with the probability of 1} ,
\label{ch_rulef1}
\EA
and the rule in case of $W_{(i\rightarrow j)}<0$
\BA
x_j(t) &\longrightarrow&
\left\{
\begin{array}{l}
x_i(t+dt) \quad \mbox{with the probability of 
$-W_{(i\rightarrow j)}(x_j(t),t)dt$} ,\\ 
x_j(t+dt) \quad \mbox{with the probability of 
$1+W_{(i\rightarrow j)}(x_j(t),t)dt$}
\end{array}  \right. , \nonumber \\
x_i(t) &\longrightarrow& \quad 
x_i(t+dt) \quad \mbox{with the probability of 1} .
\label{ch_rulef2}
\EA
Likewise, the rules for backward time direction
state that  in case of $W_{(i\rightarrow j)}>0$
\BA 
x_j(t) &\longrightarrow&
\left\{
\begin{array}{l}
x_i(t-dt) \quad \mbox{with the probability of 
$W_{(i\rightarrow j)}(x_j(t),t)dt$} ,\\ 
x_j(t-dt) \quad \mbox{with the probability of 
$1-W_{(i\rightarrow j)}(x_j(t),t)dt$}
\end{array}  \right. ,\nonumber \\
x_i(t) &\longrightarrow& \quad
x_i(t-dt) \quad \mbox{with the probability of 1} ,
\label{ch_ruleb1}
\EA
and in case of $W_{(i\rightarrow j)}<0$
\BA
x_i(t) &\longrightarrow&
\left\{
\begin{array}{l}
x_j(t-dt) \quad \mbox{with the probability of 
$-W_{(i\rightarrow j)}(x_i(t),t)dt$} ,\\ 
x_i(t-dt) \quad \mbox{with the probability of 
$1+W_{(i\rightarrow j)}(x_i(t),t)dt$}
\end{array}  \right., \nonumber \\
x_j(t) &\longrightarrow& \quad
x_j(t-dt) \quad \mbox{with the probability of 1} .
\label{ch_ruleb2}
\EA

According to the rules of the random jumping process above,
the behavior of each
sample path is illustrated as follows:  For forward time direction,
a sample path starts from $x_i(t_I)$, develops according to (\ref{ch_llf})
with $i$ for a while,
and when a chance comes, it changes its index from $i$ to $j$ and follows
(\ref{ch_llf}) with $j$ until the next jumping process takes place.
The jumping process from $x_i$ to 
$x_j$ is allowed and the reverse process is forbidden where
$W_{(i\rightarrow j)}>0$, and vice versa where $W_{(i\rightarrow j)}<0$.  
The jumping processes may be repeated
or may not occur, depending on the sign and magnitude
of $W_{(i\rightarrow j)}$.  Sample paths show similar behavior for backward
time direction. 

It is remarked that $x_i(t)$ is generally 
a functional of both of $dw_1(s)$ and $dw_2(s)$ ($s<t$) (or $dw_{*1}(s)$ or 
$dw_{*2}(s)$ ($s>t$)) as it may repeat jumps between $i=1$ and $i=2$
in the past (in the future).  Due to changes in the index for each sample path,
there are several types of averages which are distinguished from each other 
carefully. It is convenient to introduce notations for conditional averages.
The simple average $< \cdots >$ should be taken
over both of  $dw_1(s)$ and $dw_2(s)$ ($s<t$).  
To represent a physical average of the $i$-state at $t$,
we introduce a notation of
\BE \label{iaver}
\ll f(x(t))\gg_{\{x_i(t)\}} \equiv < f(x_i(t)) > 
\EE
where  the average on the left-handed side implies a conditional
average only over sample paths, labeled
by $i$ at $t$. This average should be expressed in terms of the probability 
distribution $P_i(x,t)$ as
\BE \label{iaver Paver}
\ll f(x(t))\gg_{\{x_i(t)\}} = \int\! dx
\, f(x) P_i(x,t).
\EE
The notation 
$\ll f(x(t))\gg_{\{x_1(t)\}\cup\{x_2(t)\}}$ owns its trivial
interpretations,
\BE \label{ch_totalaver}
\ll f(x(t))\gg_{\{x_1(t)\}\cup\{x_2(t)\}} =< f(x(t))> .
\EE
Furthermore, conditional averages with different times such as
$\ll f(x(t))\gg_{\{x_i(t+dt)\}\cap\{x_j(t)\}}$ can be introduced: This example
represents the average only over the sample paths which have the index
$j$ at $t$ and $i$ at $t+dt$.

Let us now evaluate the time derivative of the physical average
$\ll f(x(t)) \gg_{\{x_i(t)\}}$.  For forward time
direction, using appropriate conditional averages, we write down as
\BA
\frac{d}{dt}\ll f(x(t)) \gg_{\{x_i(t)\}}&=&\frac{1}{dt}\left[\ll 
f(x(t+dt))\gg_{\{x_i(t+dt)\}} -\ll f(x(t)) \gg_{\{x_i(t)\}}\right] 
\nonumber \\
&=&\frac{1}{dt}\left[\ll f(x(t+dt))-
f(x(t))\gg_{\{x_i(t+dt)\}\cap\{x_i(t)\}} \right. \nonumber \\
&&\qquad  +\ll f(x(t+dt))\gg_{\{x_i(t+dt)\}\cap\{x_j(t)\}} \nonumber \\
&&\qquad \left. -\ll f(x(t))\gg_{\{x_j(t+dt)\}\cap\{x_i(t)\}}\right].
\label{ch_bibun1}
\EA
The three terms here are manipulated as
\BA
\lefteqn{\ll f(x(t+dt))-f(x(t))\gg_{\{x_i(t+dt)\}\cap\{x_i(t)\}} }
\nonumber \\
&& =\ll \left. \frac{d f(x)}{dx} \right|_{x=x(t)}
dx(t)+ \left. \frac{1}{2}\frac{d^{2}
f(x)}{dx^{2}}\right|_{x=x(t)}
(dx(t))^{2}+o(dt^{3/2})\gg_{\{x_i(t+dt)\}\cap\{x_i(t)\}} \nonumber \\
&& =\ll \frac{d f(x)}{dx}
b_i(x(t),t)dt+\frac{d^{2}f(x)}
{dx^{2}}\frac{\hbar}{2m_i}dt
\gg_{\{x_i(t+dt)\}\cap\{x_i(t)\}} \nonumber \\
&& =\ll \frac{d f(x)}{dx}
b_i(x(t),t)dt+  \frac{d^{2}f(x)}
{dx^{2}}\frac{\hbar}{2m_i}dt
\gg_{\{x_i(t)\}}+o(dt^{2}) \nonumber \\
&& =dt\int\! dx\,\left(\frac{d f(x)}{dx} 
b_i(x,t)+\frac{d^{2}f(x)}
{dx^{2}} \frac{\hbar}{2m_i}\right) P_i(x,t)+o(dt^{2}) \nonumber \\
&&=dt\int\! dx\, f(x)\left(-\frac{\partial}{\partial
x}b_i(x,t)+\frac{\hbar}{2m_i}\frac{\partial^{2}}{\partial
x^{2}}\right)P_i(x,t)+o(dt^{2}),
\EA
\BE
\ll f(x(t+dt))\gg_{\{x_i(t+dt)\}\cap\{x_j(t)\}}
=- dt\int\! dx\,f(x) W_{(i\rightarrow j)}(x,t)
P_i (x,t)\theta(-W_{(i\rightarrow j)}(x,t))+o(dt^{2}),
\EE
and
\BE \label{ch_bibun2}
\ll f(x(t))\gg_{\{x_j(t+dt)\}\cap\{x_i(t)\}}
=dt\int\! dx\,  f(x)W_{(i\rightarrow j)}(x,t)
P_i(x,t)\theta(W_{(i\rightarrow j)}(x,t))+o(dt^{2}),
\EE
respectively, from (\ref{ch_llf}), (\ref{dwidwj}), (\ref{ch_rulef1}),  
(\ref{ch_rulef2}) and (\ref{iaver Paver}). 
Collecting (\ref{ch_bibun1})$\sim$(\ref{ch_bibun2}), we obtain the correct
time evolution of (\ref{ch_fokker_f}). 
This shows the equivalence between  (\ref{ch_fokker_f})
and the stochastic equation  (\ref{ch_llf}) supplemented with the 
stochastic jumping process
(\ref{ch_rulef1}) and (\ref{ch_rulef2}).  
Likewise one can show the equivalence between
(\ref{ch_fokker_b}) and the stochastic equation (\ref{ch_llb})
supplemented with the stochastic jumping process
(\ref{ch_ruleb1}) and (\ref{ch_ruleb2}).

We need some careful treatment on the dynamical condition in the 
present case. For the
equivalence between the Nelson's approach and the Schr\"odinger
approach, the dynamical condition is desired to have 
the form of
\begin{equation}\label{Newton-Nelson ch}
\frac{\partial v_i}{\partial t}
=\frac{\hbar}{2m_i}\frac{\partial^{2} u_i}{\partial x^{2}}
-v_i \frac{\partial v_i}{\partial x}+u_i \frac{\partial u_i}{\partial x}
-\frac{1}{m_i}\frac{\partial \tilde{V}_{ii}}{\partial x} .
\end{equation}
Here we introduce a ``quantum potential'' $\tilde{V}_{ii}$
which is to include the effect of channel coupling as well
as the usual potential $V_{ii}$.
The simplest way to achieve this equation is to define
the ``mean balanced acceleration" $a_i$ through
the ``mean (forward and backward) time derivative" as usual
but for the stochastic process
without any jumping process.  We simply consider a stochastic process governed
by (\ref{ch_llf}) all the time, and denote $X_i$ instead of $x_i$ 
to distinguish them from each other.  There is no mixing of $dw_i$ and $dw_j$ 
in $X_i$, contrary to $x_i$.  For each $X_i(t)$ we define the ``mean balanced 
acceleration" $a_i(X_i(t),t)$, and the ``Newton" equations,
\BE\label{X_nel_new}
m_i a_i(X_i(t),t) = - \frac{\partial \tilde{V}_{ii}}{\partial X_i},
\EE
becomes (\ref{Newton-Nelson ch}).

The combination of the equations 
(\ref{ch_kinematics})$+i$(\ref{Newton-Nelson ch}) derives
\BE
\frac{\partial}{\partial x}\left[i \frac{\hbar}{m}\frac{1}{\psi'_i}
\frac{\partial \psi'_i}{\partial t}+\frac{1}{2}\left(\frac{\hbar}{m_i}\right)^{2}
\frac{1}{\psi'_i}\frac{\partial^{2}\psi'_i}{\partial x^{2}}-\frac{1}{m_i}
\left\{\tilde{V}_{ii}-\frac{i\hbar}{2}W_{(i \rightarrow j)}\right\}\right]
=0
\label{eqpsi'}
\EE
where the relation,
\BE\label{UVch}
u_i+iv_j=\frac{\hbar}{m_i}\frac{1}{\psi'_i}\frac{\partial \psi'_i}{\partial x},
\EE
is used.  If we shift the function $\psi'_i$ to 
\BE\label{shift}
\psi_{i}(x,t)=\psi'_i(x,t)
\exp(-\frac{im_{i}}{\hbar}\int^{t}\eta(s)ds),
\EE
choose the ``quantum potential'' as
\BE
\tilde{V}_{ii}=V_{ii}+{\rm Re}\frac{\psi_{i}^{*}V_{ij}\psi_{j}}{|\psi_{i}|^{2}},
\EE
and use the relation
\BE
W_{(i \rightarrow j)}=-\frac{2}{\hbar}{\rm
Im}\frac{\psi_{i}^{*}V_{ij}\psi_{j}}{|\psi_{i}|^{2}},
\EE
we can reproduce the Schr\"{o}dinger equations (\ref{2chan}).
By the use of the (\ref{UVch}) and (\ref{shift}), the  relations,
\BA\label{b_ch}
b_i(x,t)&=&\frac{\hbar}{m_i}({\rm Im+Re})
\frac{\partial}{\partial x} \ln \psi_i(x,t) \\
\label{b*_chp}
b_{*i}(x,t)&=&\frac{\hbar}{m_i}({\rm Im-Re})\frac{\partial}{\partial x} 
\ln \psi_i(x,t) 
\EA
and (\ref{ch_rho}), are established again.

\section{Stochastic formulation for quantum system of optical potential}

In this section, let us formulate the Nelson's stochastic approach to 
a system of a single degree of freedom described by
an optical potential.
Then the Schr\"{o}dinger equation with an imaginary part
of potential, denoted by
$iU$ (a physically relevant situation, i.e., an absorptive process corresponds
to $U<0$), is written down as
\begin{equation}\label{op_schrodinger}
i\hbar \frac{\partial \psi(x,t)}{\partial
t}=\left(-\frac{\hbar^{2}}{2m}\frac{\partial^{2}}{\partial x^{2}} +V(x,t)
+iU(x,t) \right) \psi(x,t).
\end{equation}

The formulation in the previous section suggests a method to establish
a stochastic formulation for this Schr\"{o}dinger
equation.  The analogy between the channel coupling model and the 
present model becomes apparent when
we attempt the Fokker-Planck equation corresponding to 
(\ref{op_schrodinger}) in the form of 
\BA \label{op_fokker_f}
\frac{\partial P(x,t)}{\partial t}
&=& \left[-\frac{\partial}{\partial x}b+\frac{\hbar}{2m}
\frac{\partial^{2}}{\partial x^{2}}+\frac{2U}{\hbar}\right]P(x,t)
\quad \mbox{(forward in $t$)}, \\
\label{op_fokker_b}
-\frac{\partial P(x,t)}{\partial t}
&=& \left[\frac{\partial}{\partial x}b_{*}+\frac{\hbar}{2m}
\frac{\partial^{2}}{\partial x^{2}}-\frac{2U}{\hbar}\right]P(x,t)
\quad \mbox{(backward in $t$)}.
\EA
Equations (\ref{op_fokker_f}) and (\ref{op_fokker_b}) are compared 
with (\ref{ch_fokker_f}) and (\ref{ch_fokker_b}), then both are quite similar
to each other with the correspondence between  $2U/\hbar$ and
$-W_{(i\rightarrow j)}$.

While the sum of (\ref{op_fokker_f}) and (\ref{op_fokker_b})
is given by (\ref{u_}), their difference leads to
\begin{equation}\label{op_v_}
\frac{\partial P}{\partial t}=
-\frac{\partial}{\partial x}(vP) +\frac{2U}{\hbar}P
\end{equation}
instead of (\ref{v_}).
From (\ref{u_}) and (\ref{op_v_}),
follows the kinematical equation
\begin{equation}\label{op_kinematics}
\frac{\partial u}{\partial t}=
-\frac{\hbar}{2m}\frac{\partial ^{2} v}{\partial x^{2}}
-\frac{\partial}{\partial x}(uv)+\frac{1}{m}\frac{\partial}{\partial x}U,
\end{equation}
instead of (\ref{kinematics}).

The additional term in (\ref{op_fokker_f}) simply
describes production (absorption) effects for $U>0$ ($U<0$),
which one may put in such a way that 
the production (absorption) process is a transition from
a ``unphysical" sector to a ``physical" one 
(from a ``physical" sector to a ``unphysical" one).
At this point the analogy between the previous section 
and this section is helpful to find stochastic processes
equivalent to the Fokker-Planck equations 
in (\ref{op_fokker_f}) and (\ref{op_fokker_b}):  We consider
the two random variables $x_p(t)$ and $x_u(t)$ for ``physical" and
``unphysical" sectors, respectively, and stochastic jumping between them 
occurs according to certain rules, which will be specified below.  In contrast 
with the channel coupling case with the index $i$, the stochastic differential
equations for both of $x_p(t)$ and $x_u(t)$ can be common. Introducing a 
notation of a random variable
$x(t)$ standing for both of $x_p(t)$ and $x_u(t)$, we require
the same
form of stochastic differential equations for this $x(t)$ 
as (\ref{Ito_f}) and  (\ref{Ito_b}) {\em all the time},
\begin{eqnarray}
\label{llf}
dx(t)&=&b(x(t),t)dt+dw(t) \qquad \mbox{forward in time },\\
\label{llb}
dx(t)&=&b_{*}(x(t),t)dt+dw_{*}(t)
\qquad \mbox{backward in time},
\end{eqnarray}
with the same properties for $dw(t)$ as  in(\ref{dw}) and so on.
Each sample path is described by $x(t)$ as a whole, but has to be classified
into either $x_p(t)$ or $x_u(t)$ at each $t$. Typically a sample path 
changes as, for example, $x_p(t_1) \rightarrow x_u(t_2) \rightarrow \cdots
\rightarrow x_p(t_n)$ as a result of repeated jumping processes. 
A sample path is said to be physically relevant at $t$ if the sample 
is represented by $x_p(t)$, while it is not so if it is represented 
by $x_u(t)$. In other words, the physical average at
$t$ is given by the average over ensemble of not all sample paths but only
physically relevant sample paths at $t$.  The notation 
$\ll f(x(t)) \gg_{\{x_p(t)\}}$ is introduced to represent this conditional 
average for $f(x(t))$.   Similarly the notations of other conditional averages
such as $\ll f(x(t))\gg_{\{x_u(t)\}}$
and $\ll f(x(t))\gg_{\{x_p(t)\}\cup\{x_u(t)\}}$ are clear, in particular
\BE \label{op_totalaver}
\ll f(x(t))\gg_{\{x_p(t)\}\cup\{x_u(t)\}} =< f(x(t))> .
\EE
Again conditional averages related to many times can be introduced,
e.g., $\ll f(x(t))\gg_{\{x_p(t+dt)\}\cap\{x_u(t)\}}$ is supposed
to represent the average over all the sample paths
which are described by $x_u$ at $t$ and $x_p$ at $t+dt$.

Let us summarize the ``dynamical" rule for stochastic jumping processes
between $p$ and $u$. The rules are given as follows (Fig. 2):
(i) For forward time direction,
in case of $U<0$,
\BA
x_p(t) &\longrightarrow&
\left\{
\begin{array}{l}
x_u(t+dt) \quad \mbox{with the probability of 
$-2U(x_p(t),t)/\hbar dt$} ,\\ 
x_p(t+dt) \quad \mbox{with the probability of 
$1+2U(x_p(t),t)/\hbar dt$}
\end{array}  \right. , \nonumber \\
x_u(t) &\longrightarrow& \quad 
x_u(t+dt) \quad \mbox{with the probability of 1} ,
\label{op_rulef1}
\EA
and in case of $U>0$,
\BA
x_u(t) &\longrightarrow&
\left\{
\begin{array}{l}
x_p(t+dt) \quad \mbox{with the probability of 
$2U(x_u(t),t)/\hbar dt$} ,\\ 
x_u(t+dt) \quad \mbox{with the probability of 
$1-2U(x_u(t),t)/\hbar dt$}
\end{array}  \right. , \nonumber \\
x_p(t) &\longrightarrow& \quad 
x_p(t+dt) \quad \mbox{with the probability of 1} .
\label{op_rulef2}
\EA
(ii) For backward time direction, 
in case of $U<0$
\BA 
x_u(t) &\longrightarrow&
\left\{
\begin{array}{l}
x_p(t-dt) \quad \mbox{with the probability of 
$-2U(x_u(t),t)/\hbar dt$} ,\\ 
x_u(t-dt) \quad \mbox{with the probability of 
$1+ 2U(x_u(t),t)/\hbar dt$}
\end{array}  \right. ,\nonumber \\
x_p(t) &\longrightarrow& \quad
x_p(t-dt) \quad \mbox{with the probability of 1} ,
\label{op_ruleb1}
\EA
and in case of $U>0$,
\BA
x_p(t) &\longrightarrow&
\left\{
\begin{array}{l}
x_u(t-dt) \quad \mbox{with the probability of 
$2U(x_p(t),t)/\hbar dt$} ,\\ 
x_p(t-dt) \quad \mbox{with the probability of 
$1-2U(x_p(t),t)/\hbar dt$}
\end{array}  \right., \nonumber \\
x_u(t) &\longrightarrow& \quad
x_u(t-dt) \quad \mbox{with the probability of 1} .
\label{op_ruleb2}
\EA
Note that for forward time direction
the jumping process from $x_p$ to $x_u$ is allowed and the reverse process 
is forbidden where $U<0$, and vice versa where $U>0$, and that when $U$ is 
non-positive everywhere, the number of sample paths described $x_p(t)$ 
decreases and that in  $x_{u}(t)$ increases as $t$ goes,
the total number being conserved.  
Regardless of the indices of  $p$ and $u$, each sample path is a 
stochastic process described by (\ref{llf}) (or (\ref{llb}))  

To prove the equivalence between the Fokker-Planck equation (\ref{op_fokker_f})
and the stochastic differential equation (\ref{llf}) with the jumping rules
(\ref{op_rulef1}) and (\ref{op_rulef2}), we calculate, for example,
\BA  
\lefteqn{ \frac{d \ll f(x(t)) \gg_{{\{x_p(t)}\}}}{dt} } \nonumber \\
&&= \frac{1}{dt}\left[\ll f(x(t+dt))\gg_{\{x_p(t+dt)\}}
-\ll f(x(t)) \gg_{\{x_p(t)\}}\right] \nonumber \\
&& =\frac{1}{dt}\left[\ll f(x(t+dt))-
f(x(t))\gg_{\{x_p(t+dt)\}\cap\{x_p(t)\}} \right. \nonumber \\
&&\qquad +\ll f(x(t+dt))\gg_{\{x_p(t+dt)\}\cap\{x_u(t)\}} \nonumber \\
&& \qquad \left. -\ll f(x(t))\gg_{\{x_u(t+dt)\}\cap\{x_p(t)\}}\right],
\label{op_bibun1}
\EA
with
\BA
\lefteqn{
\ll f(x(t+dt))-f(x(t))\gg_{\{x_p(t+dt)\}\cap\{x_p(t)\}} } \nonumber \\
&& 
=dt\int\! dx\, f(x) \left(-\frac{\partial}{\partial
x}b(x,t)+\frac{\hbar}{2m}\frac{\partial^{2}}{\partial
x^{2}}\right)P(x,t)+o(dt^{2}),
\EA
\BE
\ll f(x(t+dt))\gg_{\{x_p(t+dt)\}\cap\{x_u(t)\}} 
=dt\int\! dx\, f(x) \frac{2U(x,t)}{\hbar}P(x,t)\theta(U(x,t))+o(dt^{2}),
\EE
and
\BE \label{op_bibun2}
\ll f(x(t))\gg_{\{x_{u}(t+dt)\}\cap\{x_{p}(t)\}}
=-dt\int\! dx\, f(x) \frac{2U(x,t)}{\hbar}P(x,t)\theta(-U(x,t))+o(dt^{2}).
\EE
These equations (\ref{op_bibun1})$\sim$(\ref{op_bibun2})
follows (\ref{op_fokker_f}).
The equivalence between the Fokker-Planck approach and the approach of 
the stochastic differential equation (\ref{llf})
with the stochastic jumping process 
(\ref{op_rulef1}) and (\ref{op_rulef2}) has 
been shown for forward direction.   
Similarly the equivalence between the two approaches can 
be proven for backward time direction.

As for the dynamical condition, we do not modify the original 
Nelson's formulation.  When the mean time derivatives $Df(t)$ and
$D_{*}f(t)$ are concerned, there may be some ambiguity with respect to
taking expectation.  Here we will follow the argument given
above (\ref{X_nel_new}). We define the ``mean balanced acceleration''
through the ``mean time derivatives'' as usual but for the
stochastic process without any jumping process.
We simply consider a stochastic process governed by (\ref{llf})
and (\ref{llb}) all time. This leads to the 
``Newton-Nelson equation" in (\ref{Newton-Nelson})
in  the present case.

The combination of the equations 
(\ref{op_kinematics})$+i$(\ref{Newton-Nelson})
leads to
\BE
\frac{\partial}{\partial x}[i \frac{\hbar}{m}\frac{1}{\psi '}
\frac{\partial \psi '}{\partial t}+\frac{1}{2}(\frac{\hbar}{m})^{2}
\frac{1}{\psi '}\frac{\partial^{2}\psi '}{\partial x^{2}}-\frac{1}{m}(V+iU)]
=0
\EE
where the relation (\ref{u+iv}) is used.
Again the relation between $\psi'$ and the solution
of (\ref{op_schrodinger}) $\psi$ is given as
\BE
\psi=\psi' \exp(-\frac{im}{\hbar}\int^{t}\eta(s) ds),
\EE
and 
\BA\label{b_op}
b(x,t)&=&\frac{\hbar}{m}({\rm Im+Re})\frac{\partial}{\partial x} \ln \psi(x,t), \\
\label{b*_op}
b_{*}(x,t)&=&\frac{\hbar}{m}({\rm Im-Re})\frac{\partial}{\partial x} \ln \psi(x,t), \\
\label{rho_op}
P(x,t)& =&|\psi(x,t)|^{2}.
\EA
are satisfied.


\section{Numerical analysis}

Now we can perform the numerical analysis of the effects of
the optical potential and channel coupling on the tunneling
time, using above generalized Nelson's approach.

First, we discuss one-dimensional system with a static
square well optical potential,
\begin{equation}
V(x)=\left\{\begin{array}{cll}
0 &\quad \mbox{in I} \quad  &(x<0),\\
V_{0}-iU_{0} &\quad \mbox{in II} \quad &(0<x<d),\\
0 &\quad \mbox{in III} \quad &(d<x)
\end{array}\right.
\end{equation}
(Fig. 3).
We set the solution of the Shr\"{o}odinger equation 
\BE
i\hbar\frac{\partial}{\partial t}\psi(x,t)=
[-\frac{\hbar^{2}}{2m}\frac{\partial^{2} }{\partial
x^{2}}+V(x)]\psi(x,t)
\EE
as
\BE \label{psi}
\psi(x,t)=
\int_{-\infty}^{\infty}
A(k)\varphi_{k}(x) e^{-i\frac{E}{\hbar}t}dk
\EE
with a coefficient function $A(k)$ and
$E=\frac{\hbar^2 k^2}{2m}$.
It is well-known that $\varphi_{k}(x)$ is written as
\begin{equation} \label{varphi}
\varphi_{k}(x)=\left\{\begin{array}{cl}
e^{ikx}+R_{k} e^{-ikx} &\quad\mbox{in I},\\
C_{k} e^{\kappa x}+D_{k} e^{-\kappa x}
&\quad\mbox{in II},\\
T_{k} e^{ikx} &\quad\mbox{in III},
\end{array}\right.
\end{equation}
where
\begin{equation}
\kappa=\frac{\sqrt{2m(V_{0}-iU_{0}-E)}}{\hbar}=\kappa_{R}-i\kappa_{I}
\quad (\kappa_{I}>0),
\end{equation}
\begin{equation}
(\kappa_{0}=\frac{\sqrt{2m(V_{0}-iU_{0}-E_{0})}}{\hbar}=\kappa_{R0}-i\kappa_{I0}
\quad (\kappa_{I0}>0),)
\end{equation}
and $R_{k}$, $T_{k}$, $C_{k}$ and $D_{k}$ are given as
\begin{equation}
\left[ \begin{array}{c}
R_{k} \\
T_{k} \\
C_{k} \\
D_{k}
\end{array}\right]=
{\cal B}
\left[\begin{array}{c}
-i(\kappa^{2}+k^{2}) {\sinh}\kappa d \\
2k\kappa e^{-ikd} \\
k(\kappa+ik) e^{-\kappa d} \\
k(\kappa-ik) e^{\kappa d} \\
\end{array}\right],
\end{equation}
\begin{equation}
{\cal B}=\frac{1}{2k\kappa {\cosh}\kappa d +i (\kappa^{2}-k^{2})
{\sinh}\kappa d}.
\end{equation}
We take a Gaussian form with its center at $k=k_{0}$, or
\begin{equation}
A(k) = A_{k_0}(k) =C \exp \left\{-\frac{(k_{0}-k)^2}{4 \sigma ^2} \right\},
\label{Ak0}
\end{equation}
with a normalization constant $C$.
We put here $\sigma=\frac{k_{0}}{100}$ and
$V_{0}=5E_{0}=(\frac{\hbar k_{0}}{2m})^{2}$.
Using this solution, we calculate numerically (\ref{llf}),
(\ref{llb}) and (\ref{op_rulef1}) $\sim$ (\ref{op_ruleb2}) .

Figure 4 shows the three typical sample paths
calculated by  (\ref{llf}) and (\ref{op_rulef1}) and  (\ref{op_rulef2}).
There is a  sample path $x(t)$  which change it's
property from ``physical'' to ``unphysical'' in the tunnel region.

Figures 5 and 6 shows  the parameter $\frac{U_{0}}{E_{0}}$ v.s.
the average of passing time $\tau_{p}$, calculated by
(\ref{llb}), (\ref{op_ruleb1}) and (\ref{op_ruleb2}). 
See the details of this ``backward time evolution method'' 
in our previous work \cite{Imafuku=1995}.
Generally, $\tau_{p}$ decrease as the $\frac{U_{0}}{E_{0}}$
become larger.  Let us estimate $\tau_{p}$ analytically on 
the W.K.B. like approximation.
If we can write the wave function in the tunnel region II as
\BE\label{approximation}
\psi(x,t)\sim\psi(x)\sim C' \exp (- \kappa_{0} x)= 
\exp \{- (\kappa_{R0}-i\kappa_{I0}) x\},
\EE
the drift of (\ref{llb}) becomes 
\BE
b_{*}=\frac{\hbar}{m}(\kappa_{I0}+\kappa_{R0})
\sim\frac{\hbar\overline{\kappa}_{0}}{m}
(1+\frac{\kappa_{I0}}{\kappa_{R0}}
+o^{2}(\frac{\kappa_{I0}}{\kappa_{R0}})), \qquad
\overline{\kappa}_{0}= \frac{\sqrt{2m(V_{0}-E_{0})}}{\hbar}.
\EE
from (\ref{b*_op}). 
In these cases, the ``backward'' time evolution of the
distribution function $P_{T}(x,t)$, which has an ``initial''
distribution $\delta (x-d)$, is written as
\BE\label{lim_fokker}
\frac{\partial P_{T}(x,t)}{\partial t}=
-[\frac{\hbar}{m}\overline{\kappa_{0}}(1+\frac{\kappa_{I0}}{\kappa_{R0}})\frac{\partial}{\partial x}
+\frac{\hbar}{2m}\frac{\partial^{2}}{\partial x^{2}}-\frac{2}{\hbar}U_{0}]P_{T}(x,t),
\EE
and we can get the solution of (\ref{lim_fokker}) easily
\begin{equation}
P_{T}(x,t)=
\sqrt{\frac{2m\pi}{-\hbar t}}\exp\{\frac{(x-d+\frac{\hbar
\overline{\kappa}_{0}(1+\frac{\kappa_{I0}}{\kappa_{R0}})}{m}
t)^{2}}{\frac{\hbar t}{2m}}-\frac{2U_{0}}{\hbar}t\}, \qquad (t<0).
\end{equation}
There are two characteristic time intervals in this solution.
One is the diffusion time $t_{d}\sim \frac{m d^{2}}{\hbar}$
for which the distribution sizes up to the potential width
$d$.  The other is the current time
\BE \label {tcc}
t_{c}\sim \frac{m d}{\hbar \overline{\kappa}_{0}(1+\frac{\kappa_{I0}}{\kappa_{R0}})}
\sim \frac{m d}{\hbar
\overline{\kappa}_{0}}(1-\frac{\kappa_{I0}}{\kappa_{R0}})
\EE
for which the peak of the distribution moves from $x=d$ to $x=0$.
Of course, the approximation of (\ref{approximation}) is
justified when $\overline{\kappa}_{0} d$ is much larger than
$1$, and this leads us to the relation of
\begin{equation}
t_{d} \qquad \gg \qquad t_{c},
\end{equation}
and the time interval $t_{c}$ becomes the passing time
in this extreme case. Note that this $t_{c}$ has  tendency of
decreasing as the $\frac{U_{0}}{E_{0}}$ become larger.

Second, we discuss a one-dimensional system
with a static square well potential and  2-cannel coupling,
or the case of the Shr\"{o}dinger equation for this problem
written down as
\BE \label{ch_schrodinger1}
i \frac{\partial}{\partial t}
\left[\begin{array}{c}
\psi_{1}\\
\psi_{2}
\end{array} \right]
=\left[
\begin{array}{cc}
-\frac{1}{2m} \frac{\partial^2}{\partial x^2}+V
& U \\
U&-\frac{1}{2m} \frac{\partial^2}{\partial x^2}+V
\end{array}
\right]
\left[\begin{array}{c}
\psi_{1}\\
\psi_{2}
\end{array} \right].
\EE
$V$ and $U$ are supposed to
\begin{equation}
V(x)=\left\{\begin{array}{cll}
0 &\quad \mbox{in I} \quad  &(x<0),\\
V_{0}&\quad \mbox{in II} \quad &(0<x<d),\\
0 &\quad \mbox{in III} \quad &(d<x),
\end{array}\right.
\end{equation}
and
\begin{equation}
U(x)=\left\{\begin{array}{cll}
0 &\quad \mbox{in I} \quad  &(x<0),\\
U_{0}&\quad \mbox{in II} \quad &(0<x<d),\\
0 &\quad \mbox{in III} \quad &(d<x).
\end{array}\right.
\end{equation}
Figure 7 shows the schematical illustration of our simulation. 
We can diagonalize (\ref{ch_schrodinger1})  as
\BE \label{ch_schrodinger2}
i \frac{\partial}{\partial t}
\left[\begin{array}{c}
\psi_{+}\\
\psi_{-}
\end{array} \right]
=\left[
\begin{array}{cc}
-\frac{1}{2m} \frac{\partial^2}{\partial x^2}+V+U
& 0 \\
0&-\frac{1}{2m} \frac{\partial^2}{\partial x^2}+V-U
\end{array}
\right]
\left[\begin{array}{c}
\psi_{+}\\
\psi_{-}
\end{array} \right].
\EE
where
\BE\label{diago}
\psi_{+}=\frac{1}{\sqrt{2}}(\psi_{1}+\psi_{2}) \qquad
\mbox{and}\qquad
\psi_{-}=\frac{1}{\sqrt{2}}(\psi_{1}-\psi_{2}),
\EE
and we can write down the time-dependent solution of $\psi_{+}$ and $\psi_{-}$
easily as same as (\ref{psi}) or
\begin{equation} \label{psi+-}
\psi_{\pm}(x,t)=
\int_{-\infty}^{\infty}
A(k)\varphi_{\pm k}(x) e^{-i\frac{E}{\hbar}t}dk
\end{equation}
with a Gaussian coefficient function $A(k)$ and $E=\frac{\hbar^2 k^2}{2m}.$
$\varphi_{\pm k}(x)$ is (\ref{varphi}) substituted $\kappa$ for
\BE
\kappa_{\pm}=\frac{\sqrt{2m (V_{0}\pm U_{0}-E)}}{\hbar}.
\EE

Figure 8 shows the some typical sample paths
calculated by  (\ref{ch_llf}), (\ref{ch_rulef1}) and (\ref{ch_rulef2}).
There is a path  which  changes its index from $1$ to $2$ in the
passage through the tunneling region.
$t_{i=1,2}$  Figs. 9 and 10 are the averages of
the passing times over the sample paths which belong to
$\{x_i(t)\}$ at $t\rightarrow \infty$.
We can see in Fig. 10 that there is a critical parameter value of
$\frac{V_{0}-U_{0}}{E_{0}}=1$ in the behavior of $t_{1}$ and $t_{2}$.
This is understood as following;
In the case of $\frac{V_{0}-U_{0}}{E_{0}}>1$, 
the ``$-$'' channel, which is dominant in the tunnel region
II in comparison with``$+$'' one, is not tunneling channel and it
describes a particle which goes 
over the potential. Regardless of $x_1(t)$ and $x_2(t)$,
the time spent in the ``potential region'' is expected
to be agree with the one which is expected from classical
mechanics or
\BE
\frac{md}{\hbar k_{-0}},\quad\mbox{where}\quad
k_{-0}=\frac{\sqrt{2m(E_{0}-V_{0}+U_{0})}}{\hbar}. 
\EE
This is seen in the Fig. 9 too.
On the other hand, in the case of $\frac{V_{0}-U_{0}}{E_{0}}<1$,
we can approximate the wave functions $\psi_{+}$ and
$\psi_{-}$ in the thick tunnel region to
\BE
\left[\begin{array}{c}\label{init_cond2}
\psi_{+}\\
\psi_{-}
\end{array} \right]
\sim
\left[\begin{array}{c}
 0\\
C_{-}\exp (-\kappa_{-0} \ x)
\end{array} \right],
\EE
and  $\psi_{1}$ and $\psi_{2}$ to
\BE
\left[\begin{array}{c}
\psi_{1}\\
\psi_{2}
\end{array} \right]
\sim
\frac{1}{\sqrt{2}}
\left[\begin{array}{c}
C_{-}\exp (-\kappa_{-0} \ x)\\
-C_{-}\exp (-\kappa_{-0} \ x)
\end{array} \right],
\EE
where
\BE
\kappa_{-0}=\frac{\sqrt{2m(V_{0}-U_{0}-E_{0})}}{\hbar}.
\EE
So we can estimate ``passing time'' of both channels ($1$ and
$2$) at 
\BE
\frac{md}{\hbar \kappa_{-0}}
\EE
 likewise ({\ref{tcc}}).

\section{Summary and comments}

In this paper, we have analyzed the effects of inelastic
scattering on the tunneling time theoretically,
using generalized Nelson's
quantum mechanics.  This generalization enable
us to describe quantum system with  optical potential and
channel couplings in a real time stochastic approach.
In this formalism, the space-time development of dynamical
variable, e.g. coordinate of particle, is described by a
definite path determined stochastically.
Each sample path has a definite form of trajectory in
space-time diagram, while a physical quantity averaged over
the ensemble of these sample paths recovers the effect of
quantum coherence. This is true even in the Young's double
slits interference experiment. Nelson's quantum mechanics
gives each definite trajectory and the ensemble of it, but
it does not
predict which path is selected when one wants to measure the
position of a particle. 
In this sense, this  ``real time stochastic process approach''
seems to give us a new  insight into quantum mechanics  beyond
Copenhagen interpretation. On the other hand,
the effects of more general cases (many body system,
environment, temperature and so on)  are  subjects in a future and
this work would be  the first step to that study.

Recent experimental data of tunneling time using the neutron
spin echo shift through the magnetic films
\cite{Achiwa=1996}
seem to agree with the simulation based on our approach \cite{Ohba=1996}
and  this study will be reported in near future.

\section*{Acknowledgement}
This work was supported by the Grant-in-Aid for Science
Research, Ministry of Education, Science and Culture, Japan
(No. 08240232, and  04244105). 	

\newpage
{\large{\bf Figure caption}}\\
\begin{description}
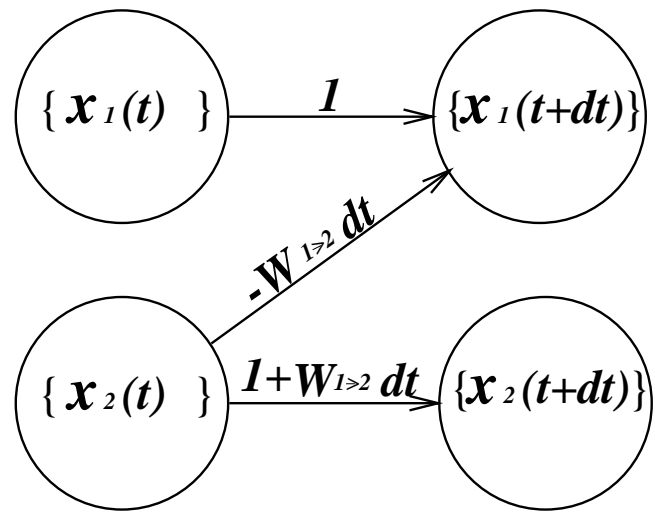
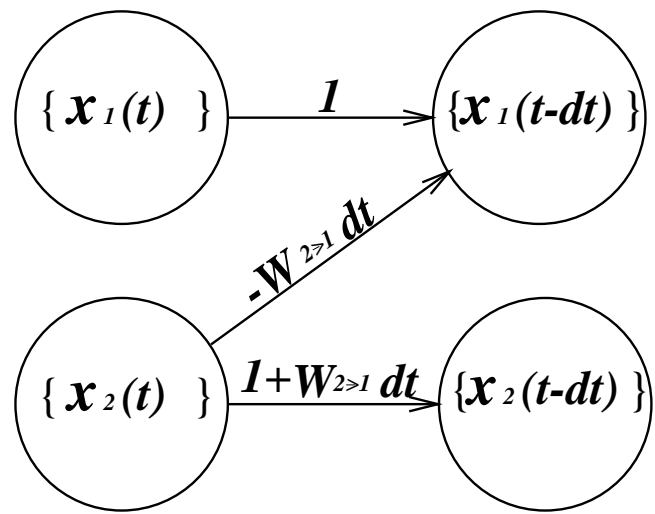
\item [FIG. 1.] Schematical illustration of the ``dynamical''
rule for stochastic jamping process between two cannels.
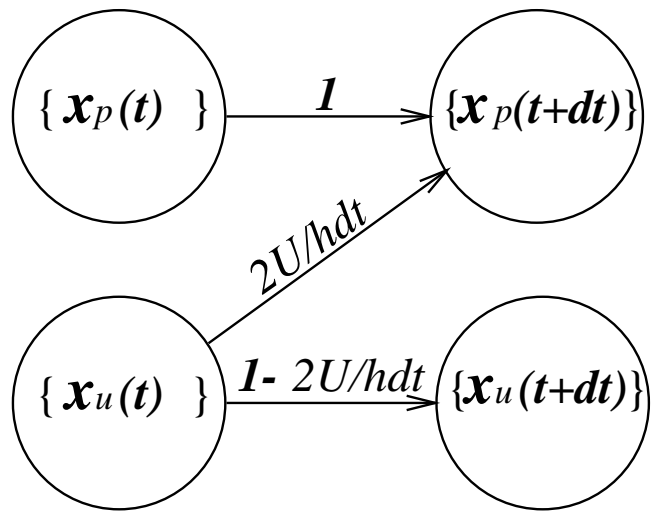
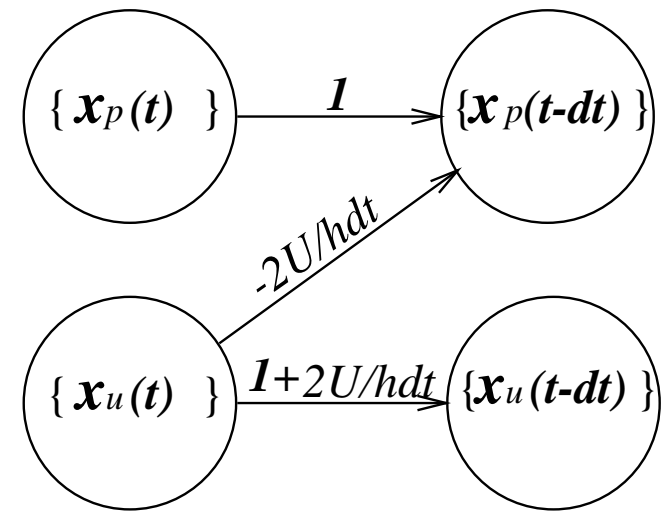
\item [FIG. 2.] Schematical illustration of ``dynamical''
rule for stochastic jamping process between physical and 
unphysical sector.
\item [FIG. 3.] Schematical illustration of one-dimensional
optical barrier tunneling.
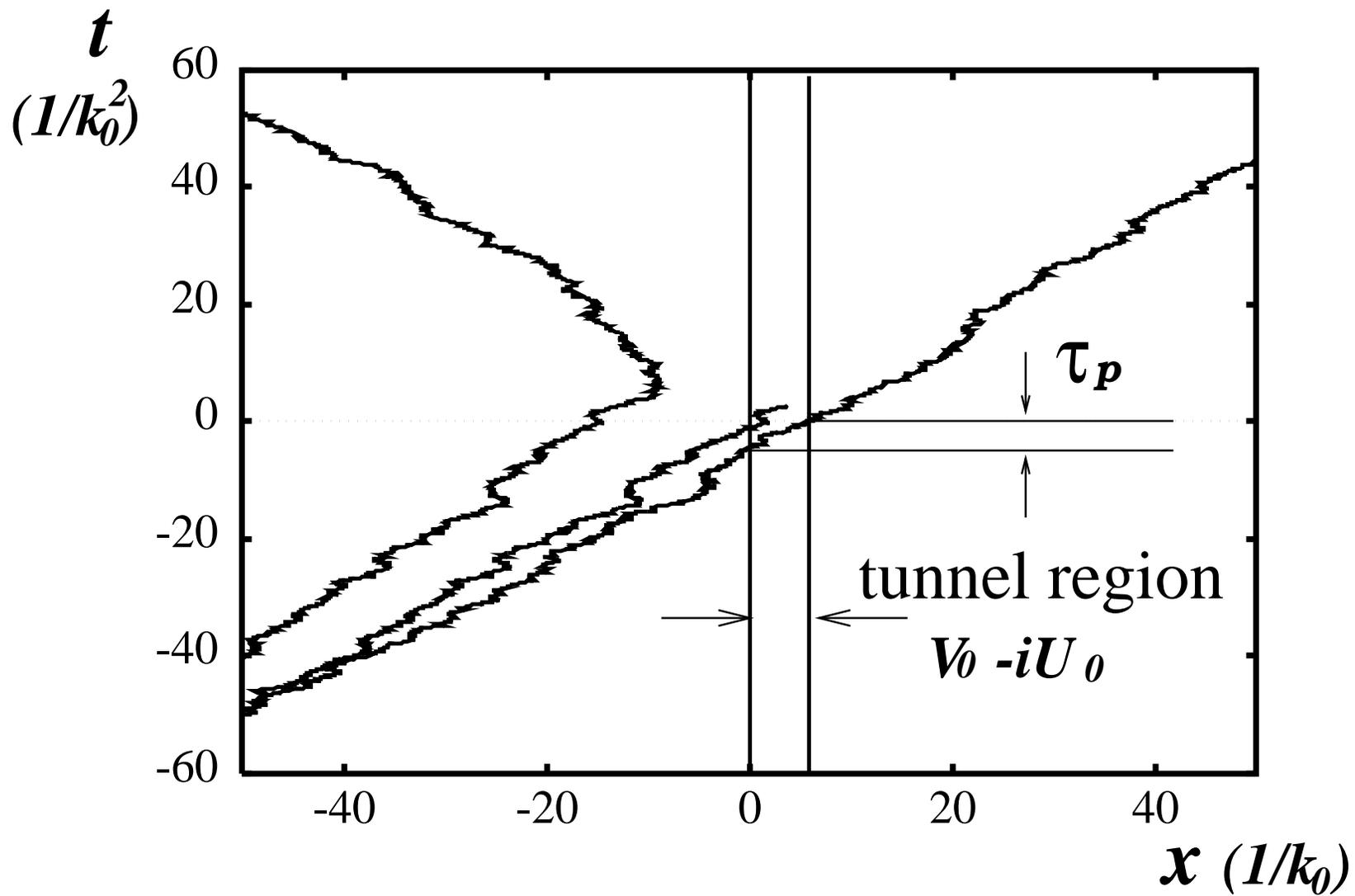
\item [FIG. 4.] The three typical sample paths in the
optical potential case.
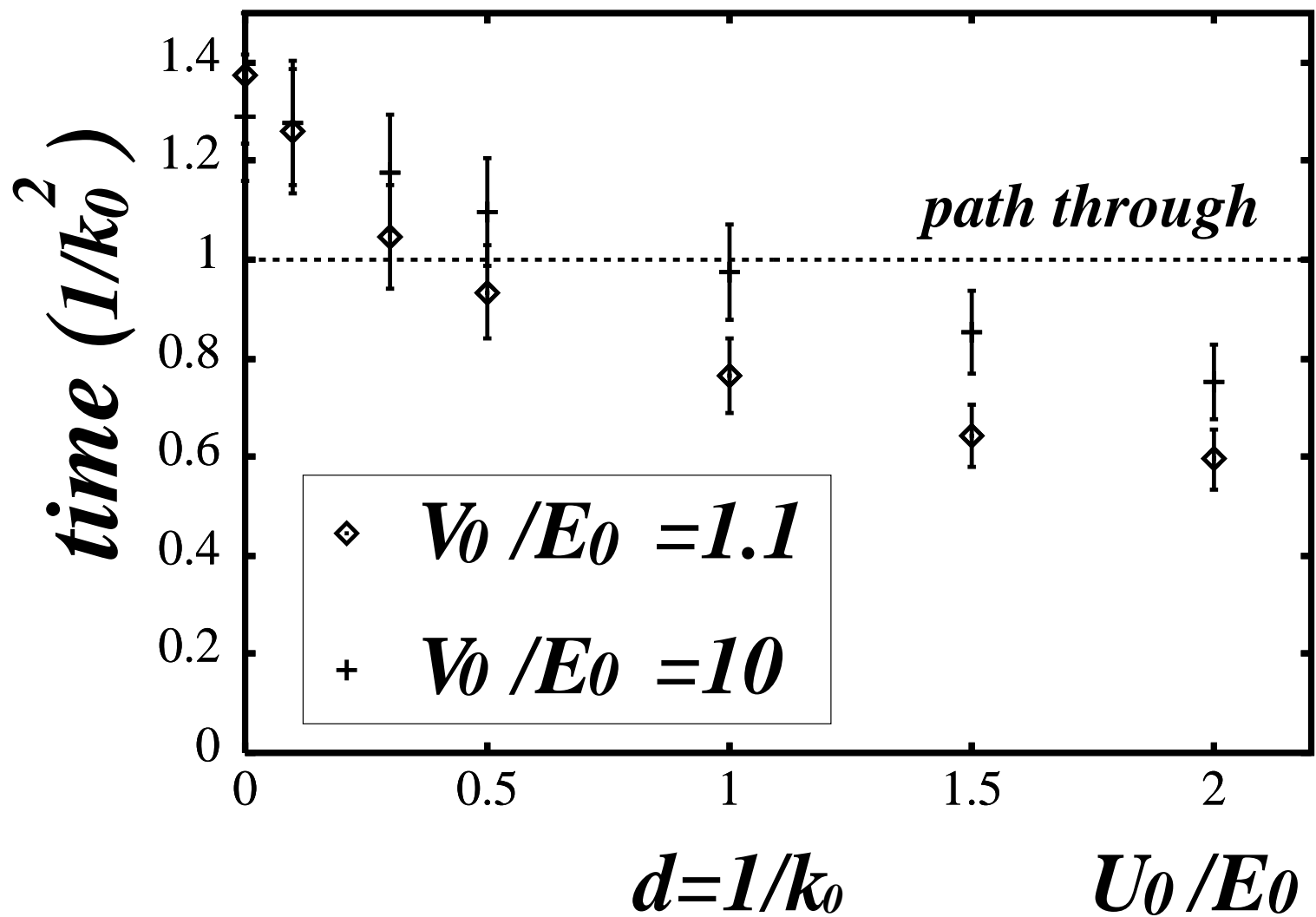
\item [FIG. 5.] The mean value of $\tau_{p}$ versus
$\frac{U_{0}}{E_{0}}$ (thin potential cases).
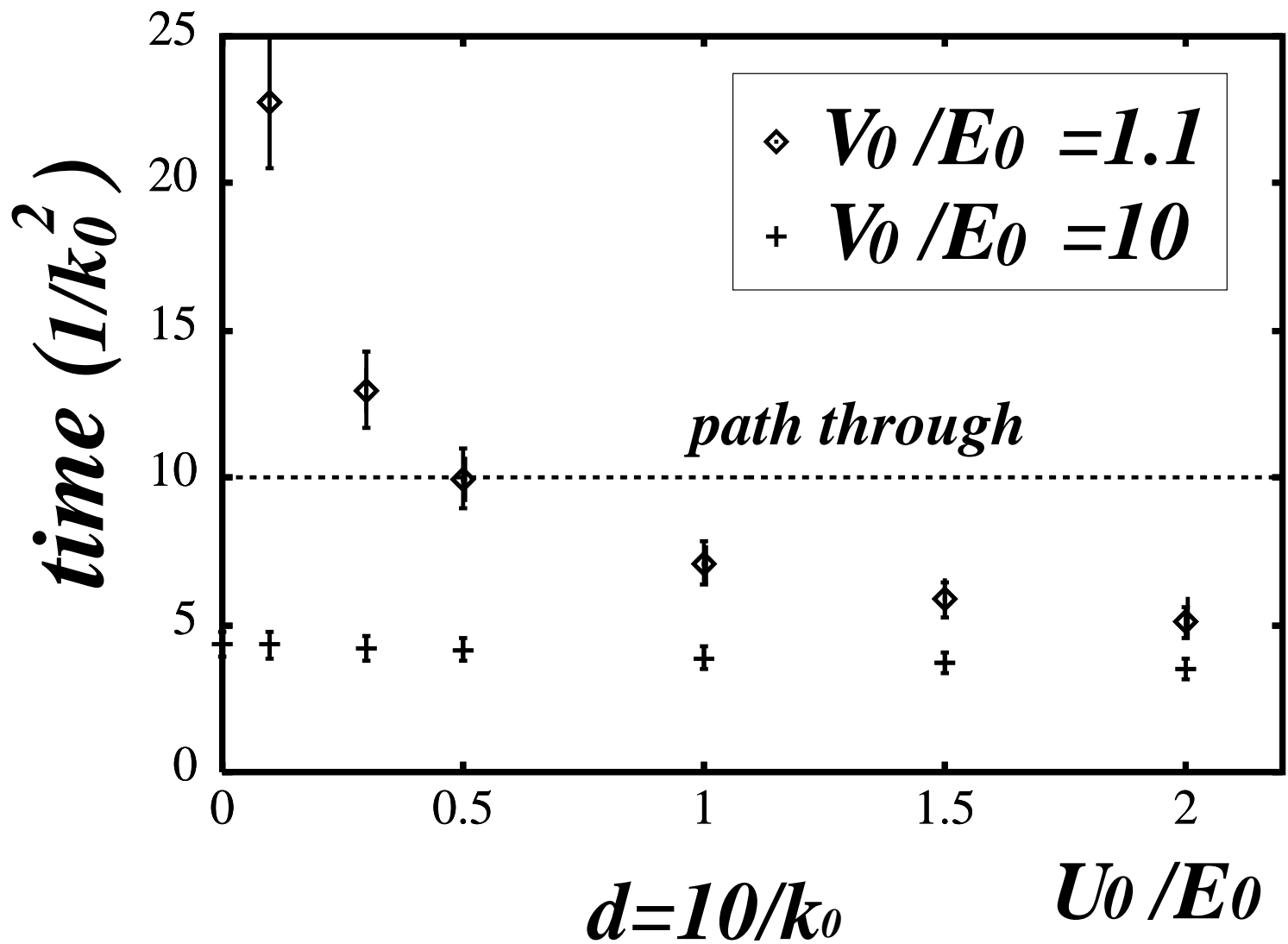
\item [FIG. 6.] The mean value of $\tau_{p}$ versus
$\frac{U_{0}}{E_{0}}$ (thick potential cases).
\item [FIG. 7.] Schematical illustration of one-dimensional
scattering with channel coupling.
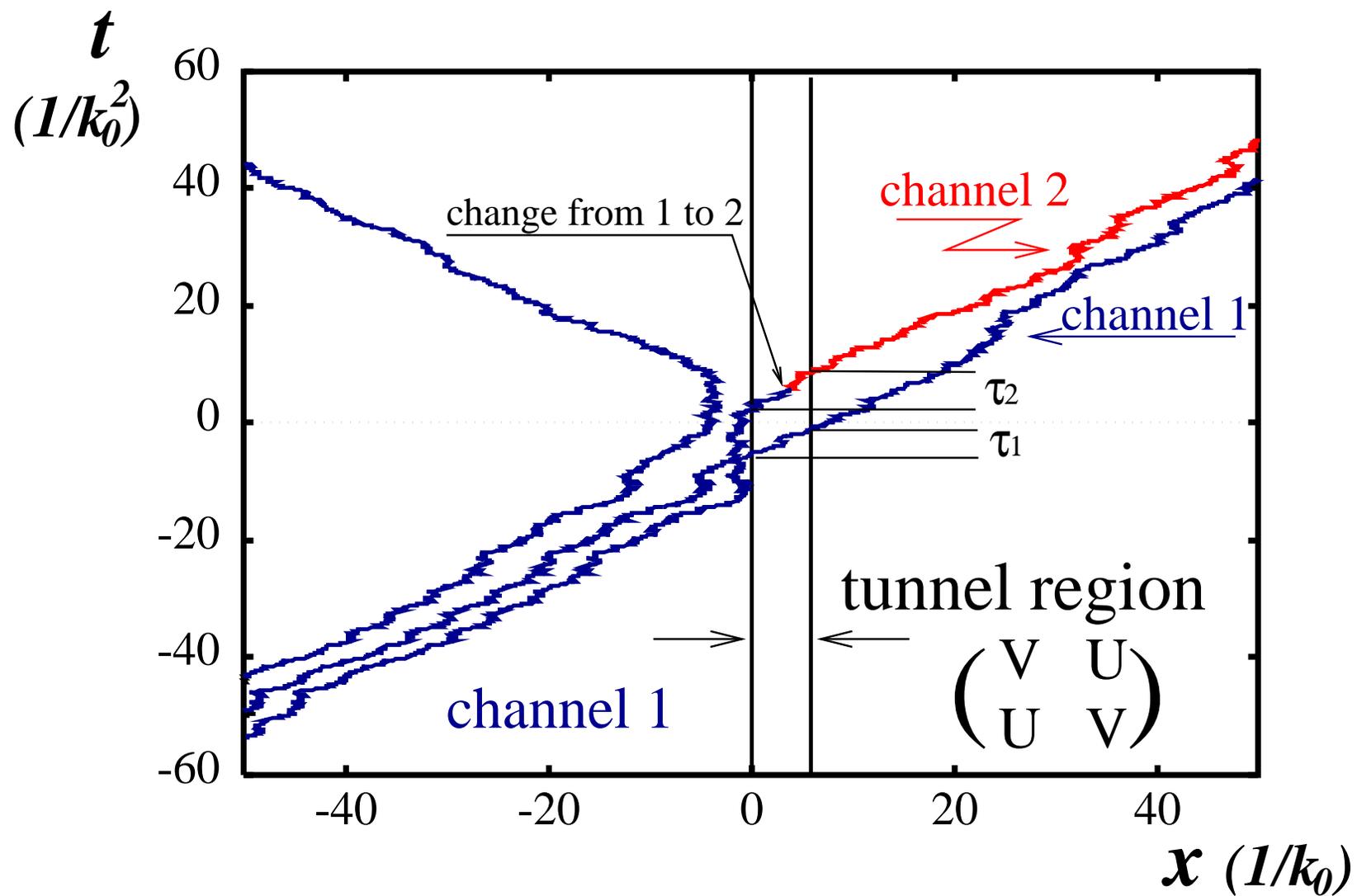
\item [FIG. 8.] The three typical sample paths with the
the channel coupling.
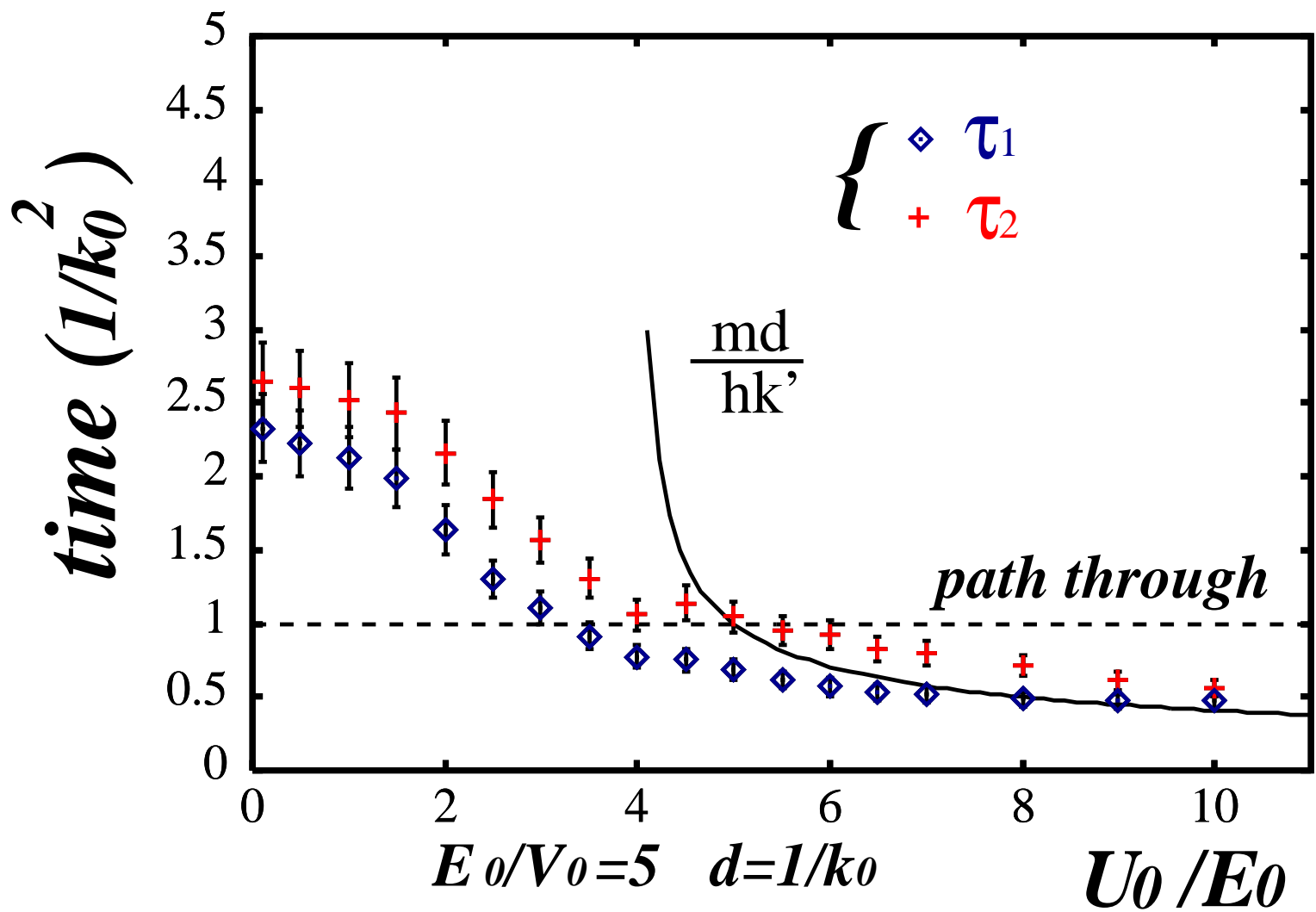
\item [FIG. 9.] The mean values of $t_{1}$ and $t_{2}$ versus
$\frac{U_{0}}{E_{0}}$ (thin potential cases).
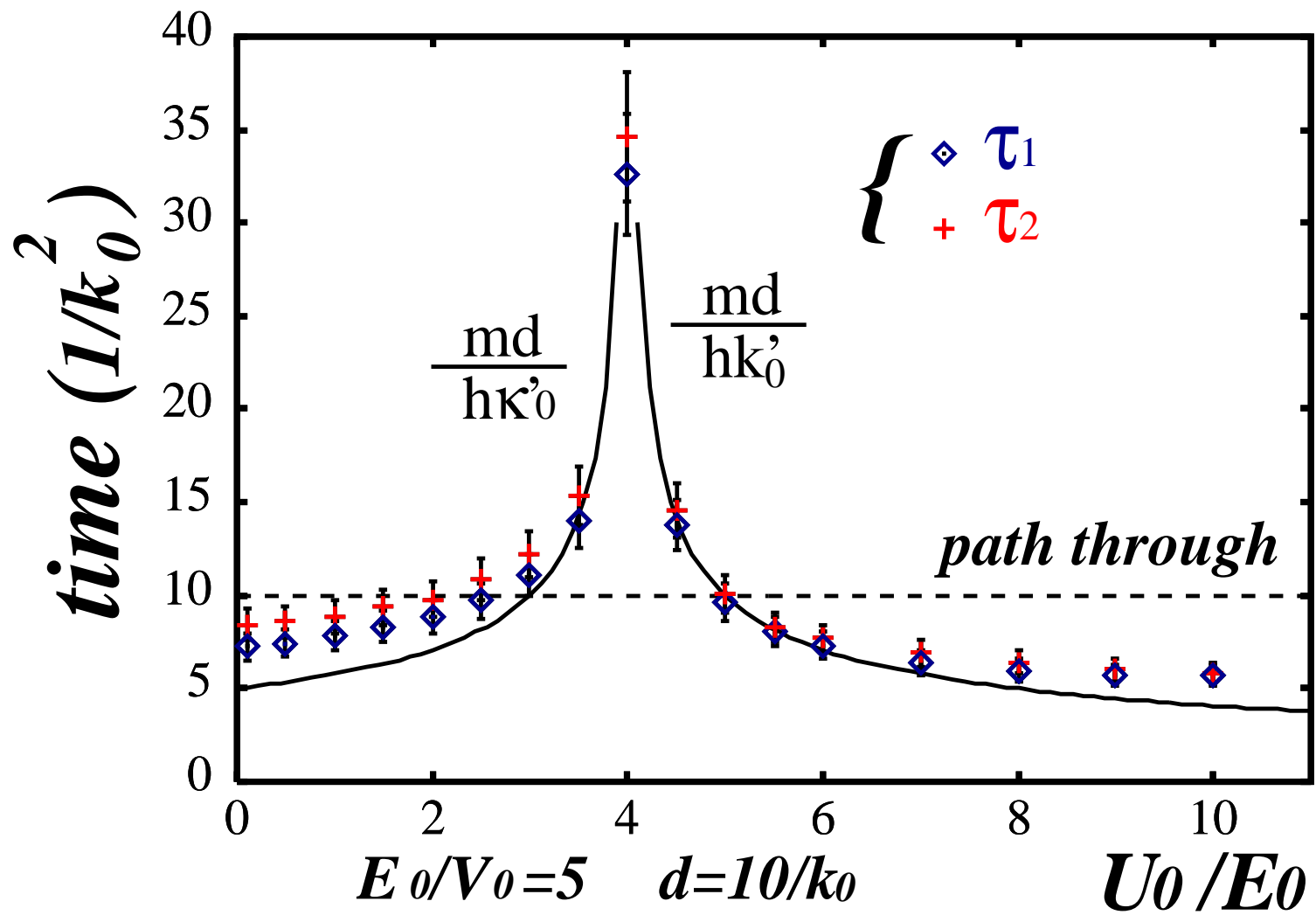
\item [FIG. 10.] The mean values of $t_{1}$ and $t_{2}$ versus
$\frac{U_{0}}{E_{0}}$ (thick potential cases).
\end{description}
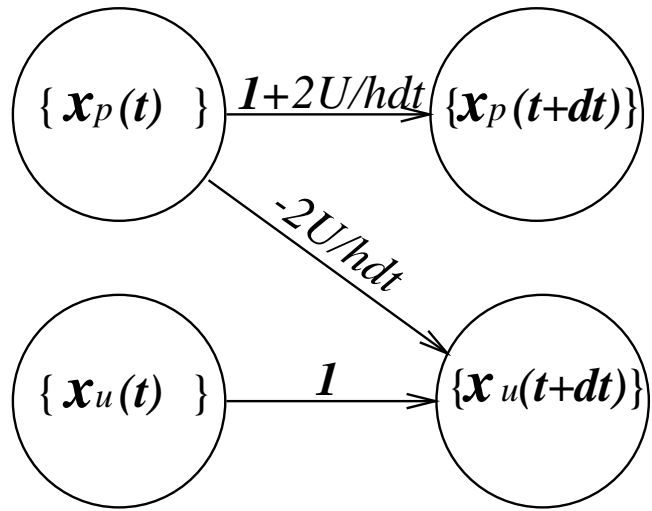
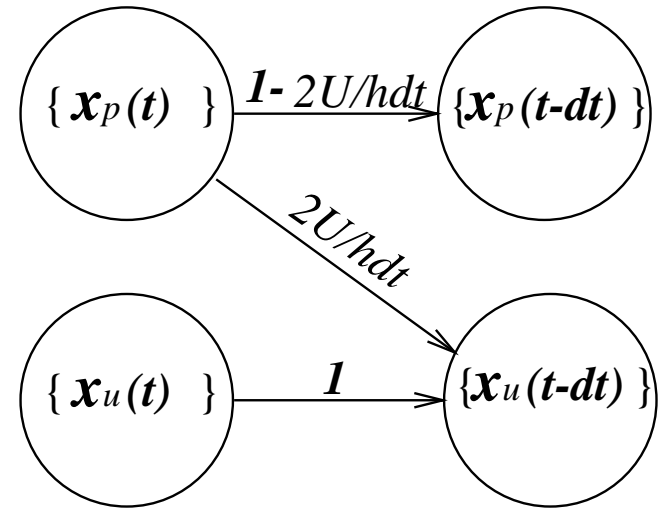
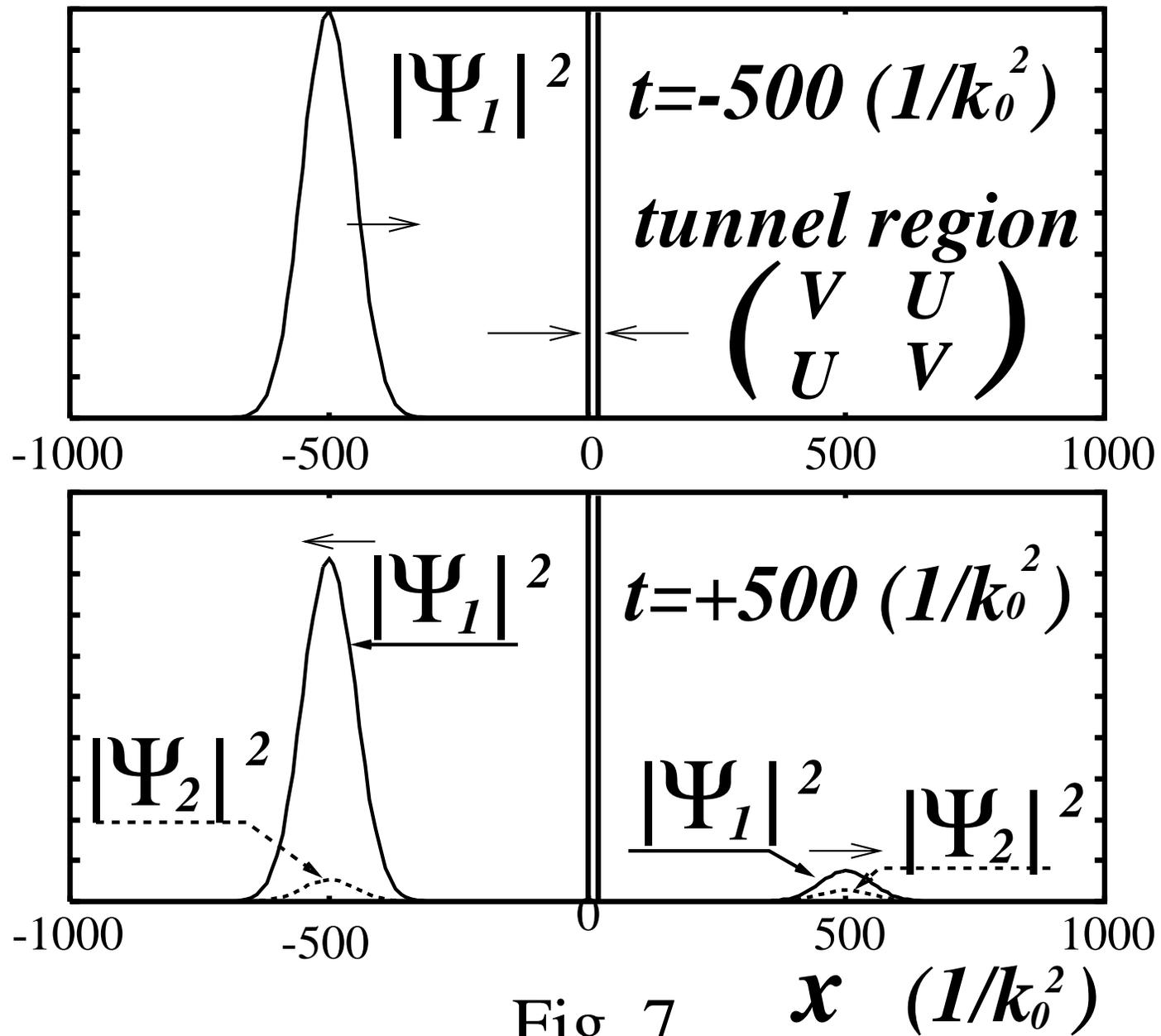
\end{document}